\documentclass[12pt]{article}

\newcommand{\Tr}{\mathop{\rm Tr}\nolimits}

\title{Construction  of Orthonormal Bases from the Vectors of the Problem
in Minkowski Space%
\thanks{Contribution to the Proceedings of the VIth International School-seminar
{\it``Actual Problems of High Energy Physics''}. Gomel, Belarus,
August 7-16, 2001}}

\author {Alexander~L.~Bondarev
\\ \sl National Scientific and Educational Center of Particle
and \\ \sl High Energy Physics of the Belarusian State University
\and \it M.Bogdanovich str.,153, Minsk, 220040, Republic of
Belarus \and \rm Tel:  (375-17) 288-34-38 \and \rm Fax: (375-17)
232-60-75 \and \rm e-mail: bondarev@hep.by}

\date{}

\begin{document}
\maketitle

\begin{abstract}
We propose to use the modified Gram -- Schmidt orthonormalization
process in Minkowski space for construction of orthonormal bases
from the vectors of the problem.
\end{abstract}


\section {Introduction}

Nowadays, methods of direct calculation of the amplitudes of
processes are intensively developed (see e.g.~\cite{r1} and
references there). In this connection the task of construction of
orthonormal bases from the vectors of the problem (in particular,
polarization bases for vector bosons) becomes actual.

In the present work, we propose to use the modified Gram --
Schmidt orthonormalization process in Minkowski space for these
purposes.

We use Feynman metrics:
\\
$ \displaystyle \mu = 0,1,2,3, \;\;\; a^{\mu} = ( a_0 , \vec{a} )
, \;\;\; a_{\mu} = ( a_0, -\vec{a} ) , \;\;\; ab = a_{\mu} b^{\mu}
= a_0 b_0 - \vec{a} \vec{b} \, $; \\
the sign of the Levi-Civita tensor is determined as
$
\displaystyle {\varepsilon}_{ 0 1 2 3 } = + 1 $.

Moreover, the generalized Gram determinants are used (see
also~\cite{r2}). For example:
$$
\begin {array}{l} 
G\pmatrix{ {}_{\mu} & {}_{\nu} \\ {}_{\alpha} & {}_{\beta} } =
\left| \matrix{  g_{\mu \alpha} & g_{\mu \beta} \\
 g_{\nu \alpha} & g_{\nu \beta} } \right| =  g_{\mu \alpha} g_{\nu
 \beta} - g_{\mu \beta} g_{\nu \alpha}
\; ,
\end {array}
g_{\mu \alpha} = \left\{
\begin{array}{rl}
            1, &  \;\; \mu = \alpha = 0  \\
           -1, &  \;\; \mu = \alpha = 1,2,3  \\
            0, &  \;\; \mu \neq \alpha
\end{array}
\right. $$
$
\begin {array}{l} \displaystyle
G\pmatrix{ a & b \\ c & {}_\beta } = G\pmatrix{ {}_{\mu} &
{}_{\nu}
\\ {}_{\alpha} & {}_{\beta} } a^{\mu} b^{\nu} c^{\alpha} =
\left| \matrix{  (a c) & a_{\beta}
\\ (b c) & b_{\beta} } \right| =  (a c) b_{\beta} -  (b c)
a_{\beta}
\end {array}
$,
and so on.


\section {Gram -- Schmidt  orthonormalization process in the $n$-dimensional
space}

We consider briefly the Gram -- Schmidt  orthonormalization
process, with the help of which one can construct an orthonormal
basis from the vectors of the problem (see e.g.~\cite{r3}).

Let a system of linearly independent vectors
$\{ x_1 , x_2 , \cdots , x_n \}$
is in the $n$-dimensional space. Then the following system of
vectors is orthogonal:
$$
 \displaystyle
y_1 = x_1  \;  ,
$$
$$ \displaystyle y_2 = x_2 - { (x_2 y_1) \over (y_1 y_1)} y_1  \;
,
$$
$$ \displaystyle y_3 = x_3 - { (x_3 y_1) \over (y_1 y_1)} y_1 - {
(x_3 y_2) \over (y_2 y_2)} y_2  \; , $$

$$
------------------------
$$
$$ \displaystyle y_n = x_n - { (x_n y_1) \over (y_1 y_1)} y_1 - {
(x_n y_2) \over (y_2 y_2)} y_2 - \cdots - { (x_n y_{n-1}) \over
(y_{n-1} y_{n-1})} y_{n-1} \; . $$
Note that%
\footnote{Here the Greek indices denote the components of vectors,
$\rho = 1,2, \cdots, n $}
$$
\begin {array}{l} \displaystyle
(y_k)_{\rho} =  { G\pmatrix{x_1 & x_2 & \cdots & x_{k-1} & x_k
\\ x_1 & x_2 & \cdots & x_{k-1} & {}_{\rho} } \over
G\pmatrix{x_l & x_2 & \cdots & x_{k-1} \\ x_1 & x_2 & \cdots &
x_{k-1} } } \;  , \; k = 1, 2, \cdots, n \; .
\end {array}
$$
The orthonormal system is formed by the vectors:
\begin {equation}
\begin {array}{l} \displaystyle
(z_k)_{\rho} = { (y_k)_{\rho} \over \sqrt{(y_k y_k)}} =  {
G\pmatrix{x_1 & x_2 & \cdots & x_{k-1} & x_k
\\ x_1 & x_2 & \cdots & x_{k-1} & {}_{\rho}  } \over
\left[ G\pmatrix{x_l & x_2 & \cdots & x_{k-1} \\ x_1 & x_2 &
\cdots & x_{k-1} } G\pmatrix{x_l & x_2 & \cdots & x_{k-1} & x_k \\
x_1 & x_2 & \cdots & x_{k-1} & x_k } \right]^{1/2} } \; .
\end {array}
\label{e2-1}
\end {equation}
%


\section {Modified orthonormalization process in Minkowski space}

In the Minkowski space, the Gram -- Schmidt orthonormalization
process has certain features.

The first, one of the vectors of the basis should be time-like.
Therefore, the 4-momentum of a massive particle can be chosen as
the first vector (further we also consider the situation where
only the massless particles participate in the reaction).

The second, because of some  symmetric Gram determinants entering
into the denominator (\ref{e2-1}) are negative (see {\bf
Appendix}), their signs should be taken into account.

Let $p$ is an arbitrary 4-momentum, such that
$ \displaystyle p^2 = m^2 \ne 0 $,
$a$, $b$ and $c$ are arbitrary vectors. Using (\ref{e2-1}), we
obtain that four vectors $l_0$, $l_1$, $l_2$, $l_3$ form an
orthonormal basis:
\begin {equation}
\displaystyle l_0 = { p \over m } \; , \label{e3-1}
\end {equation}
\begin {equation}
\begin {array}{l} \displaystyle
(l_1)_{\rho} = { G\pmatrix{p &  a \\
                             p & {}_{\rho} }
 \over m
\Biggl[ -
 G\pmatrix{p & a  \\
           p & a   }
\Biggr]^{1/2} } = { m^2 a_{\rho} - (pa) p_{\rho} \over m \sqrt{ (p
a)^2 - m^2 a^2 }  } \; ,
\end {array}
\label{e3-2}
\end {equation}
\begin {equation}
\begin {array}{l} \displaystyle
(l_2)_{\rho} = - { G\pmatrix{p & a & b \\
                           p & a & {}_{\rho}  }
 \over
\Biggl[ -
 G\pmatrix{p & a  \\
           p & a   }
 G\pmatrix{p & a & b  \\
           p & a & b   }
\Biggr]^{1/2} }
        \\[0.5cm] \displaystyle
= { \left[ a^2 (p b) - (p a) (a b) \right] p_{\rho}
  + \left[ m^2 (a b) - (p a) (p b) \right] a_{\rho}
  + \left[ (p a)^2 - m^2 a^2 \right] b_{\rho}
\over
 \sqrt{ (p a)^2 - m^2 a^2 }
 \sqrt{ 2(p a) (p b) (a b) + m^2 a^2 b^2 -
 m^2 (a b)^2 - a^2 (p b)^2 - b^2 (p a)^2 } }
\; ,
\end {array}
\label{e3-3}
\end {equation}
$$
\begin {array}{l} \displaystyle
(l_3)_{\rho}  =  { G\pmatrix{p & a & b & c\\ p & a & b & {}_{\rho}
} \over \left[- G\pmatrix{p & a & b  \\ p & a & b  } G\pmatrix{p &
a & b & c \\ p & a & b & c } \right]^{1/2} } \; .
\end {array}
$$
Finally we take into account that in the Minkowski space
$$
\begin {array}{l} 
G\pmatrix{p & a & b & c \\ p & a & b & c } = - {\varepsilon}_{
{\mu} {\nu} {\sigma} {\tau} } p^{\mu} a^{\nu} b^{\sigma} c^{\tau}
{\varepsilon}_{{\alpha} {\beta} {\lambda} {\kappa}} p^{\alpha}
a^{\beta} b^{\lambda} c^{\kappa} \; ,
\end {array}
$$
$$
\begin {array}{l} \displaystyle
G\pmatrix{p & a & b & c \\ p & a & b & {}_{\rho} } =
{\varepsilon}_{ {\mu} {\nu} {\sigma} {\tau} } p^{\mu} a^{\nu}
b^{\sigma} c^{\tau} {\varepsilon}_{ {\rho} {\alpha} {\beta}
{\lambda} } p^{\alpha} a^{\beta} b^{\lambda} \; .
\end {array}
$$
As result we have
\begin {equation}
\begin {array}{l} \displaystyle
(l_3)_{\rho} = { {\varepsilon}_{ {\rho} {\alpha} {\beta} {\lambda}
}  p^{\alpha} a^{\beta} b^{\lambda}
       \over
\Biggl[
 G\pmatrix{p & a & b  \\
           p & a & b   }
\Biggr]^{1/2} }
        \\[0.5cm] \displaystyle
= { {\varepsilon}_{ {\rho} {\alpha} {\beta} {\lambda} } p^{\alpha}
a^{\beta} b^{\lambda} \over
 \sqrt{ 2(p a) (p b) (a b) + m^2 a^2 b^2 -
 m^2 (a b)^2 - a^2 (p b)^2 - b^2 (p a)^2 } }
\; .
\end {array}
\label{e3-4}
\end {equation}
Note that in the final formulae  the vector $c$ is absent.

Thus, an orthonormal basis in the Minkowski space, using three
vectors of the problem (one of which is a 4-momentum of a massive
particle) and a total antisymmetric Levi-Civita tensor, can be
always constructed.

Construction of a basis for the reaction with massless particles
can be performed in a following way. Let $p_1$ and $p_2$ are
$4$-momentums of the problem, such that
$ \displaystyle p_1^2 = p_2^2 = 0 $.
We consider a particular form of the basis
(\ref{e3-1})~--~(\ref{e3-4}) at
$p =  p_1 +  p_2$, $m = \sqrt { 2 ( p_1 p_2 ) }$, $a = p_2$:
\begin {equation}
\displaystyle l_0 = { p_1 + p_2 \over  \sqrt{ 2 (p_1 p_2) } } \; ,
\label{e3-5}
\end {equation}
\begin {equation}
\displaystyle l_1 = { - p_1 + p_2 \over \sqrt{ 2 (p_1 p_2) } } \;
, \label{e3-6}
\end {equation}
\begin {equation}
\displaystyle (l_2)_{\rho} = { - (p_2 b) (p_1)_{\rho} - (p_1 b)
(p_2)_{\rho}  + (p_1 p_2) b_{\rho} \over
 \sqrt{ 2(p_1 p_2) (p_1 b) (p_2 b) - b^2 (p_1 p_2)^2 } }
\; , \label{e3-7}
\end {equation}
\begin {equation}
\displaystyle
(l_3)_{\rho}
= { {\varepsilon}_{ {\rho} {\alpha} {\beta} {\lambda} }
     p_1^{\alpha} p_2^{\beta} b^{\lambda}
 \over
 \sqrt{ 2(p_1 p_2) (p_1 b) (p_2 b)  - b^2 (p_1 p_2)^2 } }
\; . \label{e3-8}
\end {equation}
%


In particular when $p_1$ is the 4-momentum of a photon, vectors
\begin {equation}
\displaystyle
e^{\pm}_{\rho}(p_1, p_2, b) = - { (l_2)_{\rho} \pm i (l_3)_{\rho}
\over \sqrt{2} } = { \Tr \left[ ( 1 \pm {\gamma}_5)
{\gamma}_{\rho} {\hat p_1} {\hat b} {\hat p_2} \right]
 \over
4 \sqrt{2 (p_1 p_2) } \sqrt{ 2 (p_1 b) (p_2 b) - b^2 (p_1 p_2) } }
\label{e3-9}
\end {equation}
can be used as the polarization basis for this photon. Using the
algebra of $\gamma$-matrices, it is easy to show that
\begin {equation}
\displaystyle
{\hat e}^{\pm}(p_1, p_2, b) =  {  ( 1 \mp \gamma_5 ) {\hat p}_1
{\hat b} {\hat p}_2 + ( 1 \pm \gamma_5 ) {\hat p}_2 {\hat b} {\hat
p}_1 \over
 2 \sqrt{2 (p_1 p_2)} \sqrt{ 2(p_1 b) (p_2 b) - b^2 (p_1 p_2) } }
 \, .
\label{e3-10}
\end {equation}

Note that replacement of the vector of the problem  $b$ in
(\ref{e3-9}) by another vector $c$ leads to the appearance of the
phase factor. Really, using the identity (see e.g. \cite{r1})
$$ \displaystyle ( 1 \pm {\gamma}_5) {\hat q} Q ( 1 \pm
{\gamma}_5) {\hat q} =  \Tr \left[ ( 1 \pm {\gamma}_5) {\hat q} Q
\right] ( 1 \pm {\gamma}_5) {\hat q} \, , $$
where $q$ is an arbitrary massless vector, $Q$ is an arbitrary $4
\times 4$-matrix, we have
\begin {equation}
\displaystyle
e^{\pm}_{\rho}(p_1, p_2, b) = e^{\pm}_{\rho}(p_1, p_2, c)
e^{\displaystyle \pm i {\varphi}(p_1, p_2, b, c)} \; ,
\label{e3-11}
\end {equation}
\begin {equation}
\displaystyle
e^{\displaystyle \pm i {\varphi}(p_1, p_2, b, c)} = { \Tr \left[ (
1 \mp {\gamma}_5) {\hat p}_1 {\hat b} {\hat p}_2 {\hat c} \right]
\over 4 \sqrt{ 2 (p_1 b) (p_2 b) - b^2 (p_1 p_2) } \sqrt{ 2(p_1 c)
(p_2 c) - c^2 (p_1 p_2) } } \; .
\label{e3-12}
\end {equation}

Some details of calculations within the considered bases may be
found in \cite{r4}.

\pagebreak

\appendix{\bf \large Appendix}

\noindent
Consider the signs of symmetric Gram determinants in denominators
of the formulae (\ref{e3-1})~--~(\ref{e3-4}).
\begin{enumerate}
\item
$
\begin {array}{l} \displaystyle
 G\pmatrix{p & a  \\
           p & a   }
=  p^2 a^2 - (pa)^2
\end {array}
$,
$ p^2 = {p_0}^2 - {\vec{p} \ }^2 = m^2 $, $(p_0 > |\vec{p}| )$. \\
When $ a^2 \le 0 $, it is evident that
\begin {equation}
\begin {array}{l} \displaystyle
 G\pmatrix{p & a  \\
           p & a   } \le 0 \; .
\end {array}
\label{ea-1}
\end {equation}
Now we consider the situation when
$
 \displaystyle
a^2 = {a_0}^2 - {\vec{a} \ }^2 = {m' \ }^2 $, $(a_0 > |\vec{a}| )
$:
$$
\begin {array}{l} \displaystyle
 p^2 a^2 - (pa)^2 = ({p_0}^2 - {\vec{p} \ }^2 ) ({a_0}^2 - {\vec{a} \
 }^2) - ( p_0 a_0 - |\vec{p}| |\vec{a}| \cos{\phi} )^2
        \\ \displaystyle
 = {\vec{p} \ }^2 {\vec{a} \ }^2 (1 - {\cos}^2{\phi} ) - {p_0}^2 {\vec{a} \
 }^2 - {a_0}^2 {\vec{p} \ }^2 + 2 p_0 a_0 |\vec{p}| |\vec{a}| \cos{\phi}
            \\ \displaystyle
= {\vec{p} \ }^2 {\vec{a} \ }^2 (1 - {\cos}^2{\phi} ) - \left( p_0
|\vec{a}| - a_0 |\vec{p}| \right)^2 - 2 p_0 a_0 |\vec{p}|
|\vec{a}| (1 - \cos{\phi} )
             \\ \displaystyle
 \le {\vec{p} \ }^2 {\vec{a} \ }^2 (1 - {\cos}^2{\phi} ) - \left( p_0
|\vec{a}| - a_0 |\vec{p}| \right)^2 - 2 {\vec{p} \ }^2 {\vec{a} \
}^2 (1 - \cos{\phi} )
              \\ \displaystyle
 = - {\vec{p} \ }^2 {\vec{a } \ }^2 (1 - \cos{\phi} )^2 - \left( p_0
|\vec{a}| - a_0 |\vec{p}| \right)^2 \le 0 \;  .
\end {array}
$$
Thus, for any vector $a$,  (\ref{ea-1}) is fulfill always.

\item
By the direct calculation it can be shown that
$$\begin {array}{l}  \displaystyle
 G\pmatrix{p & a & b \\
           p & a & b  }
   \\[0.5cm] \displaystyle
=\left| \matrix{ p_0 & p_x & p_y
\\ a_0 & a_x & a_y & \\ b_0 & b_x & b_y } \right|^2
 + \left| \matrix{ p_0 & p_x & p_z
\\ a_0 & a_x & a_z & \\ b_0 & b_x & b_z } \right|^2
 + \left| \matrix{ p_0 & p_y & p_z
\\ a_0 & a_y & a_z & \\ b_0 & b_y & b_z } \right|^2
- \left| \matrix{ p_x & p_y & p_z
\\ a_x & a_y & a_z & \\ b_x & b_y & b_z } \right|^2 \; .
\end {array}$$
Under a Lorentz transformation of
$\displaystyle p^{\mu} = (p_0 , p_x , p_y , p_z)$ into $(m , 0 , 0
, 0) $,
the last term reduced to zero. Due to Gram determinant is Lorentz
invariant, we have
\begin {equation}
\begin {array}{l} \displaystyle
 G\pmatrix{p & a & b \\
           p & a & b  } \ge 0 \; .
\end {array}
\label{ea-2}
\end {equation}
\item
\begin {equation}
\begin {array}{l} \displaystyle
 G\pmatrix{p & a & b & c \\ p & a & b & c } = - \left| \matrix{
p_0 & p_x & p_y & p_z
\\ a_0 & a_x & a_y & a_z & \\ b_0 & b_x & b_y & b_z \\ c_0 & c_x
& c_y & c_z  } \right|^2 \leq 0 \; .
\end {array}
\label{ea-3}
\end {equation}
%
\end{enumerate}

Anyway, as it is well known, the equals signs in
(\ref{ea-1})~--~(\ref{ea-3}) can be only when the vectors entering
into the Gram determinants are linearly dependent.

%

\end {document}